\documentclass[figure,10pt]{jpsj2}
\usepackage{psfig,epsfig,graphics}
\def\runtitle{On the correllation effect in Peierls-Hubbard chains}
\def\runauthor{Jiri \textsc{M\'alek}, Stefan-Ludwig \textsc{Drechsler}, 
Sergej \textsc{Flach}, Eric \textsc{Jeckelmann},\\
 and Konstantin \textsc{Kladko}}
\title{On the correlation effect in Peierls-Hubbard chains}

\author{Jiri
 \textsc{M\'alek}\thanks{On leave from: Institute 
 of Physics, ASCR, Na Slovance 2, CZ-18221 Praha 8, 
 Czech Republic},  
Stefan-Ludwig \textsc{Drechsler}\thanks{drechsler@ifw-dresden.de},
Sergej \textsc{Flach}$^1$, Eric \textsc{Jeckelmann}$^2$
 and Konstantin \textsc{Kladko}$^3$}

\inst{
Leibniz-Institut f\"ur Festk\"orper- und Werkstoffforschung, D-01171 Dresden, 
PF 270116, Germany\\
$^1$
Max-Planck-Institut f\"ur Physik komplexer Systeme 
D-01187 Dresden,
 N\"othnitzer Str.\ 33, 
 Germany\\ 
$^2$ Philipps-Universit\"at Marburg, Fachbereich Physik, 
AG Vielteilchentheorie, 
D-35032 Marburg, Germany\\
$^{3}$  Ingrian Networks, 
475 Broadway Street, 
Redwood City, CA 94063, USA}

\recdate{\today}

\abst{
We reexamine  the dimerization, the charge and the spin gaps  
of a half-filled Peierls-Hubbard chain by means of the incremental 
expansion technique.
 Our numerical findings 
 are in significant quantitative conflict with recently obtained
results by M.\ Sugiura and Y.\ Suzumura 
[J.\ Phys.\ Soc.\ Jpn.\ {\bf 71} (2002) 697]
based on a bosonization and a renormalization group method,
especially with respect to the charge gap.
Their approach seems to be valid only in the weakly 
correlated case.
}

\kword{%
Peierls-Hubbard model, dimerization, charge gap, spin gap, incremental 
expansion and 
bosonization techniques, polyacetylene 
}
\begin{document}
\baselineskip 4pt
\maketitle


\section{Introduction}
The interplay of electron-electron (el-el) and electron-lattice (el-ph) 
interactions in the context of various phase transitions (instabilities) 
is one of the most challenging many-body problems in solid state physics. 
In particular, in the context of the longstanding debate how to describe 
theoretically conducting polymers like {\it trans} polyacetylene and its 
remarkable solitonic properties \cite{ssh,br}, the Peierls-Hubbard 
model (PHM) at half filling, considered as a minimal realistic model, 
became very popular \cite{br,horsch,hirsch,baeriswyl,
hayden,waas,koenig,painelli,ros,kuprievich,malek97,jeckelmann,malek98}. 
Within this 
frame in the adiabatic approximation for the lattice the behavior of the 
dimerization amplitude as a function of the two remaining model parameters, 
the onsite repulsion $U$ and the strength of the el-ph interaction is one of 
the central problems. To the best of our 
knowledge it was 
first addressed in 1981 by P.\ Horsch \cite{horsch} who predicted 
on qualitative grounds a maximal dimerization at some intermediate $U$. 
This  was confirmed later on and made more clear by approximate solutions 
obtained by 
Baeriswyl and Maki \cite{baeriswyl} and Kuprievich \cite{kuprievich} within 
the Gutzwiller approximation and the geminal techniques, respectively.
According to them the maximal dimerization occurs for weak el-ph interaction 
at $U\sim 4t_0$, where $t_0$ denotes the nearest neighbor transfer integral.
Unfortunately, the region of interest is already 
at the border line of validity of 
those approaches. 

The general nonmonotonous behavior of the dimerization amplitude 
(at weak el-ph coupling)
as a function of $U$ 
was confirmed  
by Hayden and Soos \cite{hayden} and Waas {\it et al.} \cite{waas} who 
extrapolate results of 
 exact diagonalization studies (EDS) obtained for 
 finite even-membered rings with various boundary conditions.
In the  strongly correlated limit $U\gg t_0$ the PHM can be reduced to the 
spin-Peierls model (SPM) for which a large body of literature exists
(see for instance Ref.\ \cite{otsuka} and references therein).
The general case of arbitrary $U$ and intermediate as well as strong
el-ph interactions has been covered by the present authors within the 
solitonic \cite{malek97} as well as the incremental expansion method 
(IEM) \cite{malek98} both based on self-consistent EDS.
 
In the context of the polyacetylene problem the charge gap  has been 
addressed   among others by Ro\'sciszewski and Ole\'s \cite{ros} in the 
framework of the extended PHM. There is now a general consensus that the
Peierls mechanism, i.e.\ the off-diagonal el-ph interaction, enhances the 
charge gap compared with that of the 1D-Hubbard-Mott insulator 
\cite{waas,koenig,painelli}.  
The latter problem has been solved exactly by Lieb and Wu \cite{lieb}.

Recently, the weak and intermediate region $U\leq 4t_0$ has been readdressed 
by Sugiura and Suzumura \cite{sugiura} who employ the bosonization technique 
and the renormalization group method.
Beside the dimerization they consider also the $U$ dependence of the 
charge and the spin gaps.
It is the aim of the present
 paper to show, that these calculations are in some 
conflict with results obtained for the original PHM applying the 
incremental technique supplemented by density matrix renormalization group
(DMRG) calculations\cite{jeckelmann}.

\section{Models and Methods}
Let us consider even-membered chains  with periodic boundary conditions 
modeled by 
the  Peierls-Hubbard Hamiltonian supplemented with a static harmonic 
lattice term
\begin{eqnarray}
H&=&\sum_{i,s}-t_{i,i+1}
[c^{\dag}_{i,s}c_{i+1,s}+h.c.] +\frac{1}{2}Un_{i,s}n_{i,-s} \nonumber \\
&&+\frac{K}{2}\sum_i v_i^2,
\end{eqnarray}
 where the bond alternation $v_i$ has been introduced
\begin{equation}
 \qquad v_i=u_{i+1}-u_i ,
\end{equation}
here $u_i=\pm u_0$ is the displacement of the $i$th lattice site in the dimerized state under 
consideration
and $K$ is the spring constant. We linearize the bond-length dependent 
transfer integral $t_{i,i+1}$
\begin{equation}
t_{i,i+1}=t_0-\gamma v_i .
\end{equation}
The strength of the off-diagonal 
el-ph interaction can be expressed by the dimensionless
coupling constant $g$ 
\begin{equation}
g=\gamma/\sqrt{Kt_0} <1.
\end{equation}
It is convenient to measure  the dimerization amplitude 
$v_0=2u_0$ also in dimensionless units\cite{kuprievich}
\begin{equation}
 d=v_0\sqrt{K/t_0} < 1.
\end{equation}
Values of $g \sim 0.4$ to 0.5 are typical for conducting polymers.
For the convenience of the reader we note that
our quantities $g$ and $d$ are simply related to the notations 
$\lambda \equiv \lambda_{SSH}$ introduced by Su, Schrieffer and Heeger 
\cite{ssh} as well as $t_d$ used by Sugiura and Suzumura 
\cite{sugiura} 
\begin{equation}
\lambda=4g^2/\pi , \qquad t_d=dgt_0.
\end{equation}
Thus their two values of $\lambda$ =0.25, 0.42 correspond to our
$g$=0.443113 and 0.574341 (used in the calculation shown in Figs.\ 1-4),
 respectively.
 
Let us start with the calculation of the dimerization amplitude $d$.
Naturally in the present homogeneous dimerized case 
it can be obtained from the direct minimization of the total energy per
site with respect to the single variable $d$ at fixed $U$ and $g$,  
in other words there is no need 
to solve the general self-consistency equations presented in Ref.\ 
\cite{malek97}.
According to the incremental expansion technique the ground state energy per 
site in the $n^{th}$ order reads \cite{malek98}
\begin{equation}
\epsilon^{n}(U,g,d)=\frac{1}{2}\left[ E_{2n+2,2n+2}(U,g,d)-
E_{2n,2n}(U,g,d)\right],
\end{equation}
where $E_{2n,2n}(U,g,d)$ is the total energy of a chain segment 
(finite open chain) with $2n$ electrons and $2n$ sites.
The dimerization amplitude in the infinite chain limit ($N \rightarrow \infty$,
where $N=2n+2$ is the maximal number of interacting sites  in the 
$n^{th}$ incremental expansion order)
has been extrapolated using the empirical law (see also \cite{malek97})
\begin{equation}
d(U,g,N)=d_{\infty}(U,g)+\left(\sum_{i=0} a_i(U,g)/N^i \right)\exp(-\xi (U,g)N).
\end{equation}
 Notice that $\xi$ is close to the coherence length (half-width of a 
 soliton as obtained for instance by the solitonic method\cite{malek97}
  explicitly).

 As the next step the gaps have been calculated for the  extrapolated 
 dimerization 
 $d_\infty$. 
 In the $n^{th}$ order of the incremental expansion  the charge gap 
(for definition see Ref.\ \cite{lieb}) 
reads
\begin{eqnarray}
{\Delta \cal{E}}
_{\rho}^{(n)}
\equiv \mu^+-\mu^-
=&(n+1)\left[ E_{2n+3,2n+2}-E_{2n+2,2n+2}\right] -n\left[E_{2n+1,2n}-E_{2n,2n}
\right] \nonumber \\
&-(n+1)\left[ E_{2n+2,2n+2}-E_{2n+1,2n+2}\right] -n\left[ E_{2n,2n}-E_{2n-1,2n}
 \right] \nonumber \\
=&(n+1)\left[ E_{2n+3,2n+2}-2E_{2n+2,2n+2}+E_{2n+1,2n+2} \right] \nonumber \\
&-n\left[ E_{2n+1,2n}-2E_{2n,2n}+E_{2n-1,2n}\right] ,
\label{deltachim}
\end{eqnarray}
where $\mu^{\pm}$ denotes the chemical potential for adding an electron or a 
hole, respectively.

The spin gap $\Delta {\cal{E}}_\sigma$ 
can be calculated using an analogous expressions as for $\mu^+$
(i.e.\ the first line of Eq.\ (\ref{deltachim})) 
\begin{equation}
{\Delta \cal{E}}_{s}^{(n)}=(n+1)\left( E_{2n+2,2n+2}^{T}-E_{2n+2,2n+2}^{S}\right)-
n\left( E_{2n,2n}^{T}-E_{2n,2n}^{S}\right),
\label{deltagap}
\end{equation}
where $E_{T(S)}$ is the total energy of the lowest state
in the $S_z=1 (S_z=0)$ subspace. These states correspond usually  
 to the 
triplet and singlet 
states, respectively. 
For a more detailed derivation of Eqs.\ (9,10) and technical details of the incremental 
technique 
 the reader is 
referred to Refs.\ 
\cite{malek98,malek02}.

Notice that in the expression for the charge gap (Eq.\ (9))
the corresponding multiplet notation (doublet or singlet) has been omitted
for the sake of shortness. Let's note also that in 
 Ref.\ \cite{sugiura} the charge and spin gaps have been 
denoted by $\Delta_{\rho}$ and $\Delta_{\sigma}$, respectively. 

The Peierls contribution to the total charge 
gap as a function of $U$ has been determined using the relation
\begin{equation}
\delta {\cal{E}}^{(n)}_{\rho}(U,g)={\Delta \cal{E}}^{(n)}_{\rho}(U,g)-
{\Delta \cal{E}}^{(n)}_{\rho}(U,0),
\end{equation}
where ${\Delta \cal{E}}^{(n)}_{\rho}(U,0)$ denotes the charge gap 
of the undimerized chain calculated in the $n^{th}$ incremental order.
The quantity
$\delta {\cal{E}}^{(n)}_{\rho}(U,g)$ has then extrapolated by a polynomial fit
with the leading term $\propto N^{-2}$
to the infinite chain limit $\delta {\cal{E}}_{\rho}(U,g)$. 
 Finally, $\delta {\cal{E}}_{\rho}(U,g)$ has been 
added to the charge gap   of the 
equidistant 
undimerized chain
(sometimes called also Mott-Hubbard gap) 
given by the well-known Lieb-Wu solution \cite{lieb}
\begin{equation}
{\Delta \cal{E}}_{\rho}(U,0)=U-4-8\sum_{n=1}^\infty (-1)^n\left[ 
\sqrt{\left(1+(nU/2)^2\right)}-nU/2\right]. 
\end{equation}
This way we circumvent some difficulties with the size dependence of the total 
charge gap which stem mainly from the slow 
size dependence of the Mott-Hubbard gap at small $U$ \cite{nishimoto}.

\section{Results and discussion}
Let us start with the discussion of our numerical results obtained 
up to 6$^{th}$ order in the incremental expansion using the method of exact 
diagonalization for the Peierls-Hubbard Hamiltonian of the chain segments.
All results have been extrapolated as mentioned in the previous section. 
 Dimerization amplitudes $t_d$ 
are shown in
Fig.\ 1. 
From the comparison with Fig.\ 4 of Ref.\ \cite{sugiura} one realizes 
reasonable
agreement in the weak coupling region below 1.5 $U/t_0$ 
but a nearly 100 \% larger dimerization amplitude 
at $U/t_0$=4 in our calculation.
This is mainly caused by the shift of the maximum position 
$U_{max}[g]$ to higher $U$-values and to a more pronounced asymmetric shape 
of the 
whole curve (see Fig.\ 1, right panel).
Notice that in 
Fig.\ 4 of Ref.\ \cite{sugiura} 
there is  essentially no shift of the maximum dimerization towards 
smaller $U$-values
with increasing el-ph interaction $\lambda$ as it should be.
In the weak coupling limit it has to tend 
to $U_{max}(0) \approx 3t_0$ whereas  
$U_{max}$ tends to zero for $g \rightarrow g_c\approx$ 0.69.
This is in contrast with the statement of $U_{max}$=2.2 $\pm 0.1$ for 0.2$\leq 
\lambda \leq$ 0.5 made in Ref.\ \cite{sugiura}.
 Furthermore, the difference
$\delta d=d(U_{max})-d(0)$ {\it decreases} with increasing $g$ contrarily with 
the results shown in Fig.\ 4 of Ref.\ \cite{sugiura} where it increases by a 
factor of two. 
The same problem is also evident from the inspection of Fig.\ 3 of Ref.\ 
\cite{sugiura}. For $\lambda $ approaching the critical value $\lambda_c
\approx $0.606189  
mentioned above, the $t_d(U=0)$ curve should intersect all $t_d$-curves 
at finite $U$, 
because for all el-ph constants $\lambda > \lambda_c$
 the maximal dimerization is already 
reached in the uncorrelated limit $U$=0.
(Strictly speaking, their figure Fig.\ 3 
shows only $\lambda \leq $0.6, i.e.\ slightly below $\lambda_c$. Anyhow, no 
visible change in the slope of the $t_d(\lambda)$-curves with finite $U$ 
approaching the closely lying critical point  $\lambda_c$
can be detected.)

The charge and the spin gaps are shown 
in Figs.\ 2,3.  
One readily realizes (i) that $\delta {\cal{E}}_{\rho}(U,g)$ is always 
positive, 
i.e.\ the Peierls 
mechanism {\it enhances} the total charge gap and (ii) it follows closely 
the spin gap (see Fig.\ 3). Roughly speaking, there are three regions: the 
weakly coupled  one $U/t_0\leq $1.5, the intermediately 
coupled and the strongly coupled ones. In the first one correlation effects can be 
neglected and the charge gap being almost coinciding with the spin gap 
is dominated by the el-ph interaction. In the second intermediate region  
 the Peierls mechanism 
yields a non-negligible contribution to the total charge gap.
 This is of 
interest for conducting polymers, where $U\sim U_{max} \sim 3t_0$. Here this 
Peierls contribution amounts  about
 1/3 of the total charge gap, whereas in the strongly coupled 
spin-Peierls region it becomes increasingly small reflecting this way the 
asymptotical decoupling of spin and charge degrees of freedom.
Quite interestingly, the dimerization, the Peierls-contribution to the charge 
gap as well as the spin gap exhibit a very similar dependence on $U$. 
From this observation one might 
suggest that all those three quantities are mainly described by the spin 
degrees 
of freedom (see also Ref.\ \cite{malek97} where an effective spin-Peierls 
model has been proposed to describe the dimerization $d(U,g)$ even analytically).

From our calculations it 
becomes clear that
 there is {\it no} visible hump in the charge gap curve 
near the region where
the dimerization amplitude $t_d(d)$ and the spin gap reach their maximum. 
More strikingly, in Ref.\ \cite{sugiura} the calculated
charge gap at $U=4t_0$  is even {\it smaller} than the 
Lieb-Wu predictions (see Eq.\ (12)).
This would imply a {\it negative} 
impact of the Peierls mechanism in sharp contrast 
with our findings (see Fig.\ 3).

Finally, below we report also on some incremental expansion results where due 
to
the application of the DMRG-method 
to open chain segments 
rather high orders could be considered. Furthermore,  
charge gaps obtained from total energy differences of 
long open chains (denoted as standard open chain 
results) will be provided, 
where 
the well-known definition of the charge gap \cite{lieb} 
\begin{equation}
\Delta {\cal{E}}_{\rho}(U,g,N)=E(+e,U,g)-2E(0,U,g)+
E(-e,U,g,N)
\end{equation}
has been used. 
Here
$E(0,U,g,N)$ and $E(\pm e,U,g,N)$ denotes the total energy of the neutral and 
charged open chain, respectively, at fixed $d$.
 
For instance, for $g=0.4$ and $U/t_0=4$ in the 31$^{st}$ incremental 
order one arrives 
at $d$ =0.10591 (0.10590).
With that dimerization we obtained in the 63$^{rd}$ order the charge gap 
 $\Delta {\cal{E}}_{\rho } / t_0$=1.4992 (1.5195)
whereas the standard open chain result for $N$=128 sites 
is $\Delta {\cal{E}}_{\rho}/t_0$=1.5077.
Since the charge gap of the undimerized chain   
calculated from Eq.\ (12)  equals to 1.2867 one realizes a significant 
enhancement of the charge gap due to the finite dimerization.
Here and below, the numbers in brackets denote the results obtained 
by the extrapolation of the 
incremental data calculated up to the 6$^{th}$ order.  
At the same incremental orders and open chain lengths 
as mentioned above, for $g=0.45$ (which is very close 
to $\lambda=0.25$ considered in Ref.\ \cite{sugiura}), and 
 for $U/t_0$=2 we found 
$d=$ 0.13575 (0.13650) and $\Delta {\cal{E}}_{\rho}/t_0$=0.5208 (0.5127)
whereas the standard open chain 
result is $\Delta {\cal{E}}_{\rho}/t_0$=0.5374
(compare with  0.1728 from Eq.\ (12)).
For $g=0.5$ and the same $U/t_0$=2 as above we have now  
$d=$ 0.18578 (0.18580) and $\Delta {\cal{E}}_{\rho}/t_0$=0.6696 (0.6764)
whereas the standard open chain result is $\Delta {\cal{E}}_{\rho}/t_0$=0.6828. 
For the same  $g=0.5$ but $U/t_0$=7, one arrives at $d$=0.11977  
and $\Delta {\cal{E}}_{\rho}/t_0$=3.9734 (3.9925).
The standard open chain result is $\Delta {\cal{E}}_{\rho}/t_0$=3.9777 
whereas Eq.\ (12) yields 3.7720, only.

Thus the open chain results (Eq.\ (13)) 
provide an {\it upper} bound whereas the incremental 
technique (Eq.\ (9)) 
results in a {\it lower} bound for the charge gap. The incremental 
technique 
yields excellent results for the gaps 
in the intermediate and strongly correlated regime 
whereas in the weakly correlated regime 
slight deviations occur, if one restricts oneself to sixth order dictated 
by the computational facilities available for exact diagonalization at present.
Due to an exponential size dependence 
in the dimerization amplitude (see Eq.\ (8))  it gives excellent results 
even here. 
Anyhow, if the incremental technique is combined with the DMRG technique, 
the higher  orders 
necessary 
for the weakly correlated cases can be calculated with high enough 
accuracy.
In other words, the incremental expansion technique combined with 
DMRG  method provides a very powerful tool to evaluate with high accuracy 
the dimerization amplitude, 
the charge gap as well as the spin gap.   

To conclude, from the critical comparison of our discrete calculations
with the bosonization (continuum) approach proposed by Sugiura and Suzumura 
\cite{sugiura} one has the impression that the proposed formalism is useful
only in the weak correlation region $0 \leq U/t_0 \leq $1.5. It might be 
interesting to elucidate to what extent such an approach might be improved
for instance in the spirit of the finite-band approach proposed by 
Gammel \cite{gamel} to improve the shortcomings of the continuum models 
\cite{brazovskii,lin}
compared with the noninteracting discrete SSH-model \cite{ssh}. 
Alternatively, an improved and extended effective 
spin-Peierls model based on the description proposed in 
Ref.\ \cite{malek97} would be noteworthy, too, at least in the intermediately
 strong correlated case where the maximum of the dimerization occurs for 
 weak el-ph interaction.
\\

{\bf Acknowledgement}

We thank V.\ Janis, H.\ Eschrig, and P.\ Fulde for 
helpful discussions. 
This work was supported by the  DFG project Es85/7-1
as well as by the Grant Agency of the Czech Republic 
(Project 202/01/0764; J.M.). 


\newpage
\noindent
{\Large FIGURE CAPTIONS}\\

\vspace{1cm}
\noindent
Fig.1\\
Alternating part $t_d=gdt_0$ (dimerization) of 
the transfer 
integral $t$ vs.\ on-site (Hubbard) repulsion $U$ for the
two el-ph coupling strengths  considered in Ref.\ \cite{sugiura} 
(left panel). The same 
as in the left panel for  
a larger scale of $U$ and slightly changed el-ph coupling constants $g$ 
(right panel).\\
\vspace{0.5cm}

\noindent
Fig.\ 2\\
The Peierls contribution to the total charge gap (see 
Eqs.\ (8,10) and the spin gap (Eq.\ (9) vs.\ on-site repulsion $U$. For comparison  the charge gap of the 
equidistant infinite chain, i.e.\ the Mott-Hubbard gap (full line, no symbols) 
is shown too.\\

\vspace{0.5cm}
\noindent
Fig.\ 3\\
The Peierls contribution to the total charge gap and 
the spin gap vs.\ on-site repulsion $U$.\\

\vspace{0.5cm}
\noindent
Fig.\ 4\\
Comparison of extrapolated and 6$^{th}$-order 
incremental calculations.\\
\vspace{1cm}
\begin{figure}[h]
\begin{center}
\begin{minipage}{14.2cm}
\begin{center}
\begin{minipage}{7cm}
\hspace{-3.8cm}
\psfig{file=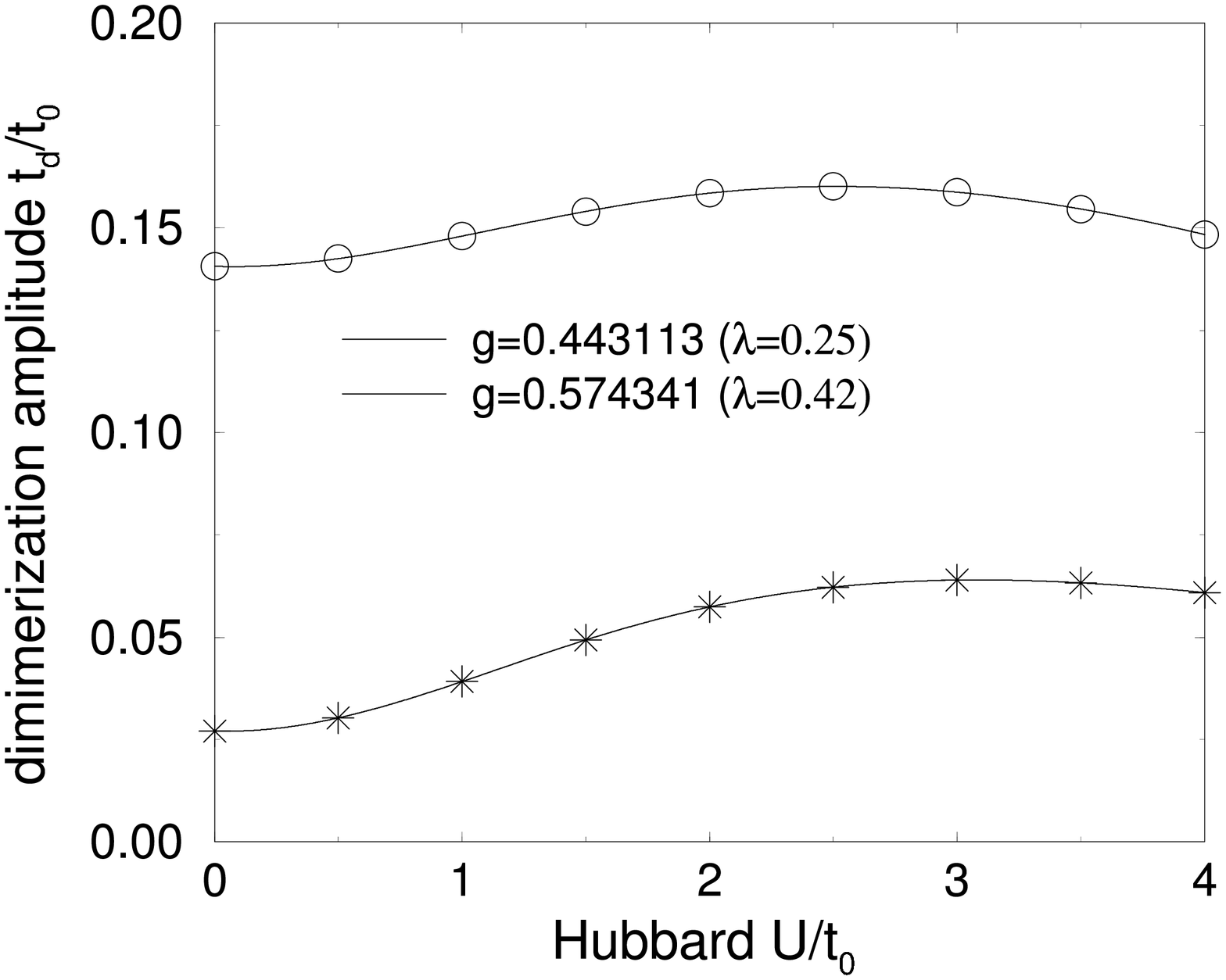,angle=-90,width=6.cm}
\hspace{2.5cm}
 \psfig{file=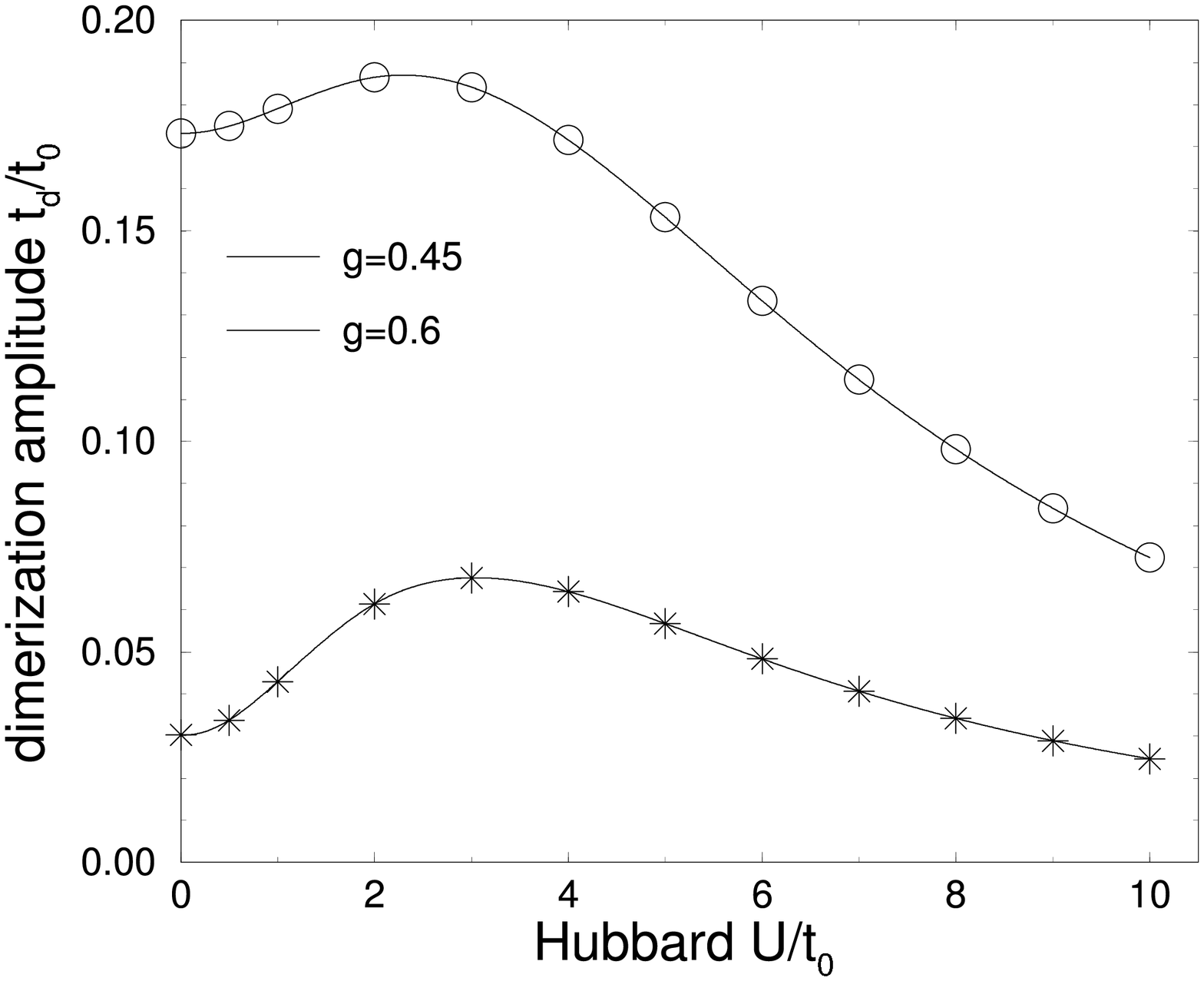,angle=-90,width=6.cm}
\end{minipage}
\end{center}
\vspace{0.cm}
\caption{}
\label{fig1}
\end{minipage}
\end{center}
\end{figure}

\begin{figure}[h]
\begin{center}
\begin{minipage}{14.2cm}
\begin{center}
\begin{minipage}{7cm}
\hspace{-3.8cm}
\psfig{file=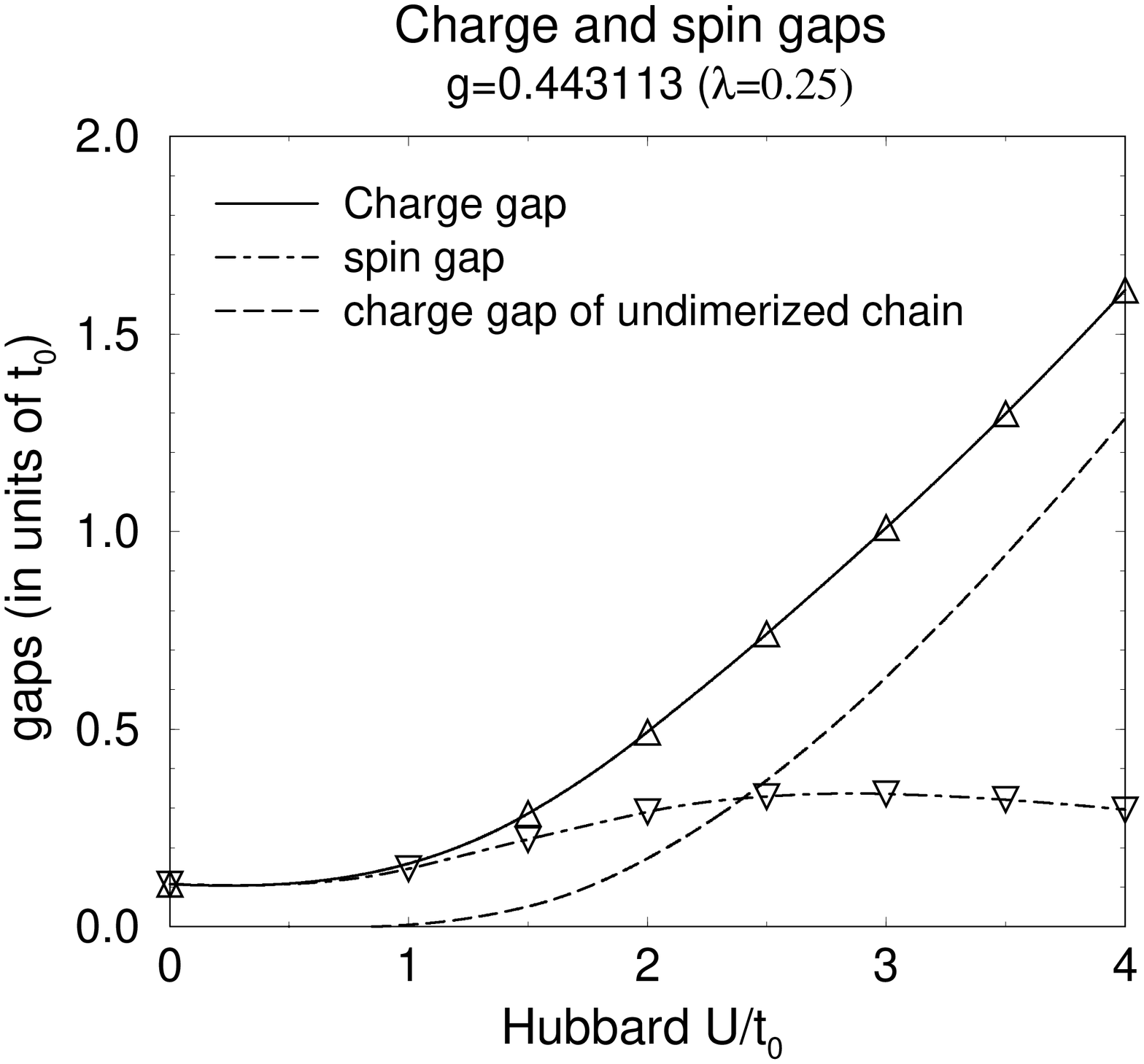,angle=-90,width=6.0cm}
\hspace{2.5cm}
\psfig{file=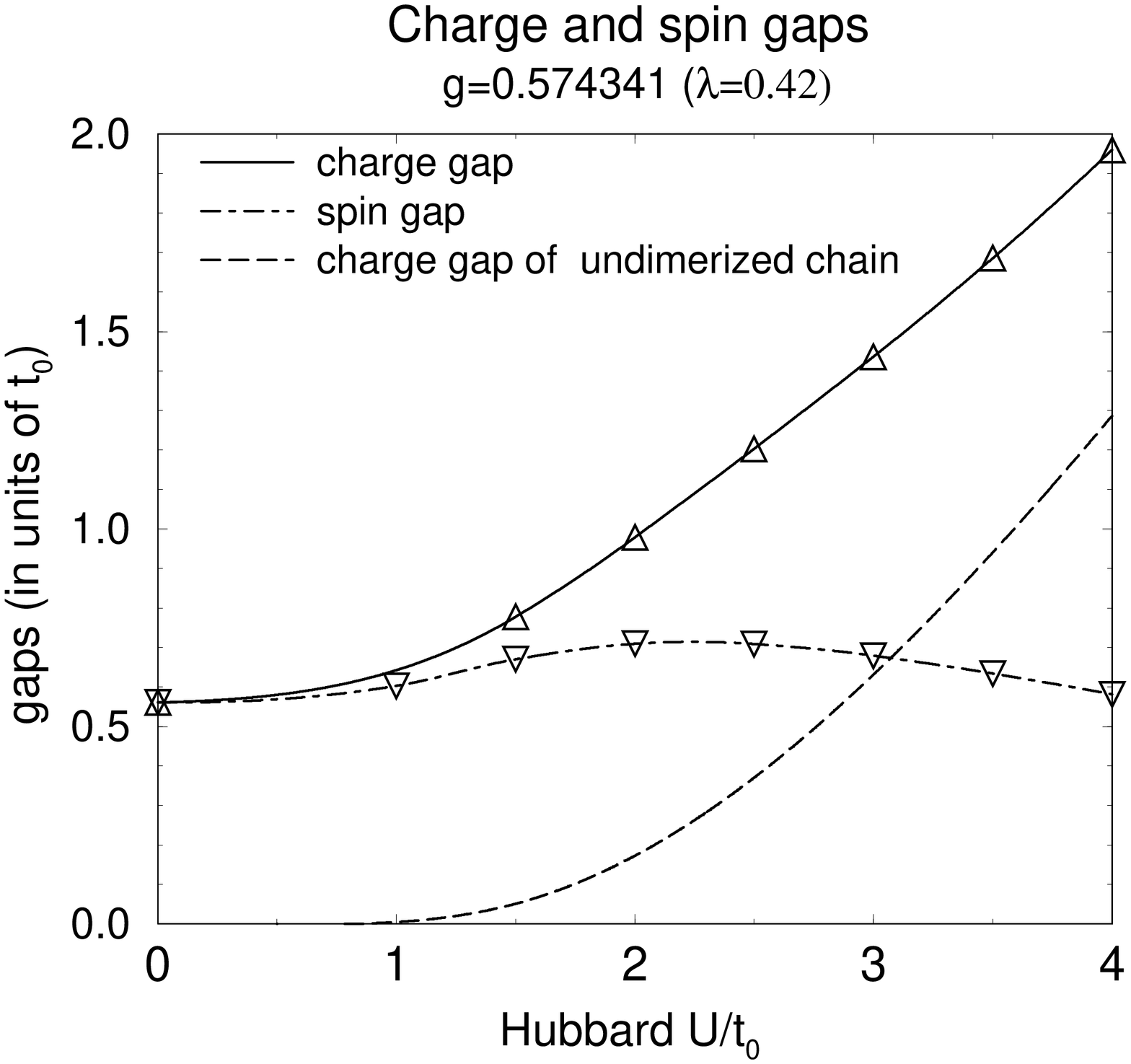,angle=-90,width=6.0cm}
\end{minipage}
\end{center}
\vspace{-0.0cm}
\caption{}
\label{fig3}
\end{minipage}
\end{center}
\end{figure}

\begin{figure}[h]
\begin{center}
\begin{minipage}{14.2cm}
\begin{center}
\begin{minipage}{7cm}
\psfig{file=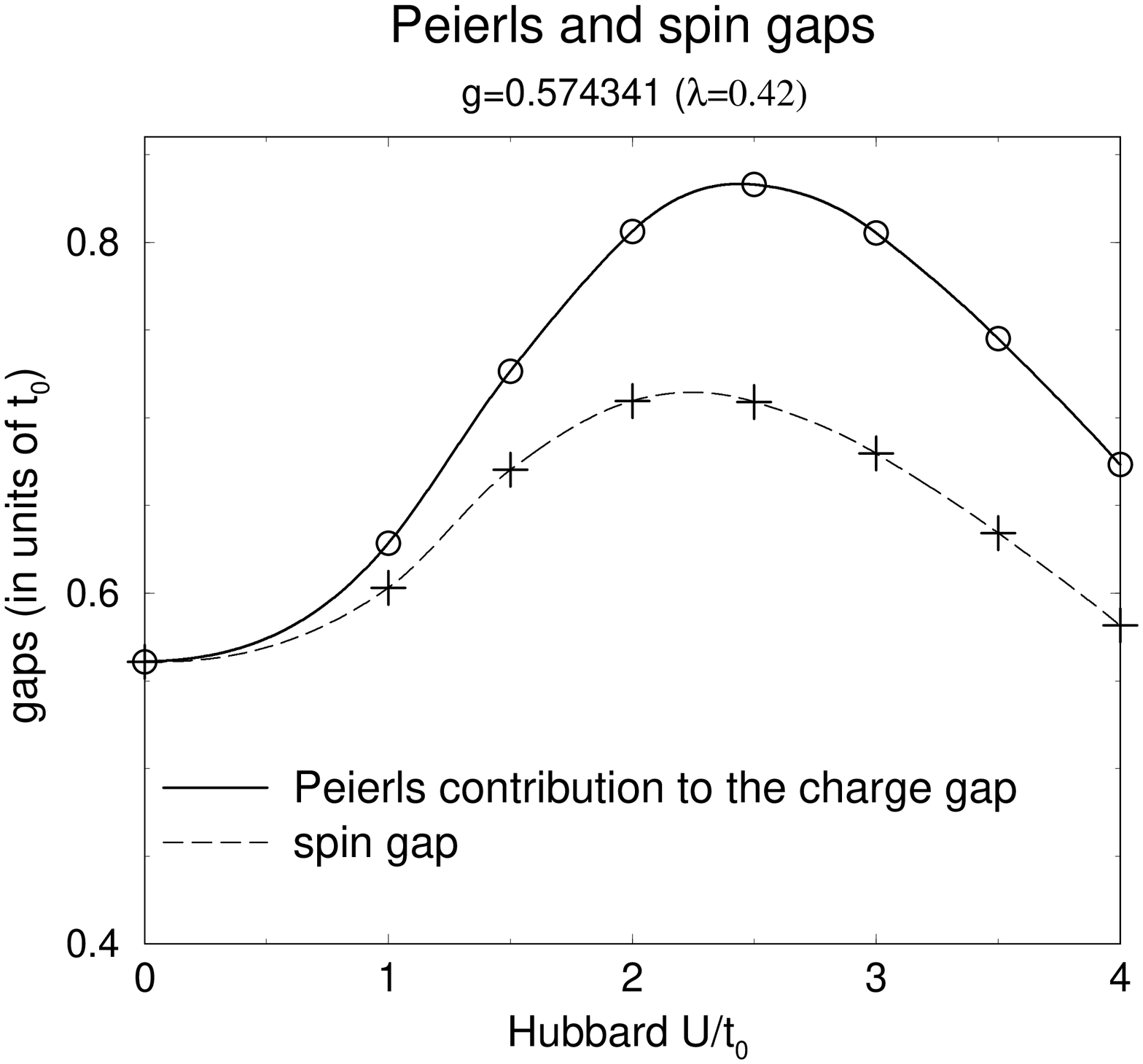,angle=-90,width=7.0cm}
\end{minipage}
\end{center}
\vspace{-0.0cm}
\caption{} 
\label{fig4}
\end{minipage}
\end{center}
\end{figure}
\begin{figure}[h]
\begin{center}
\begin{minipage}{14.2cm}
\begin{center}
\begin{minipage}{7cm}
\psfig{file=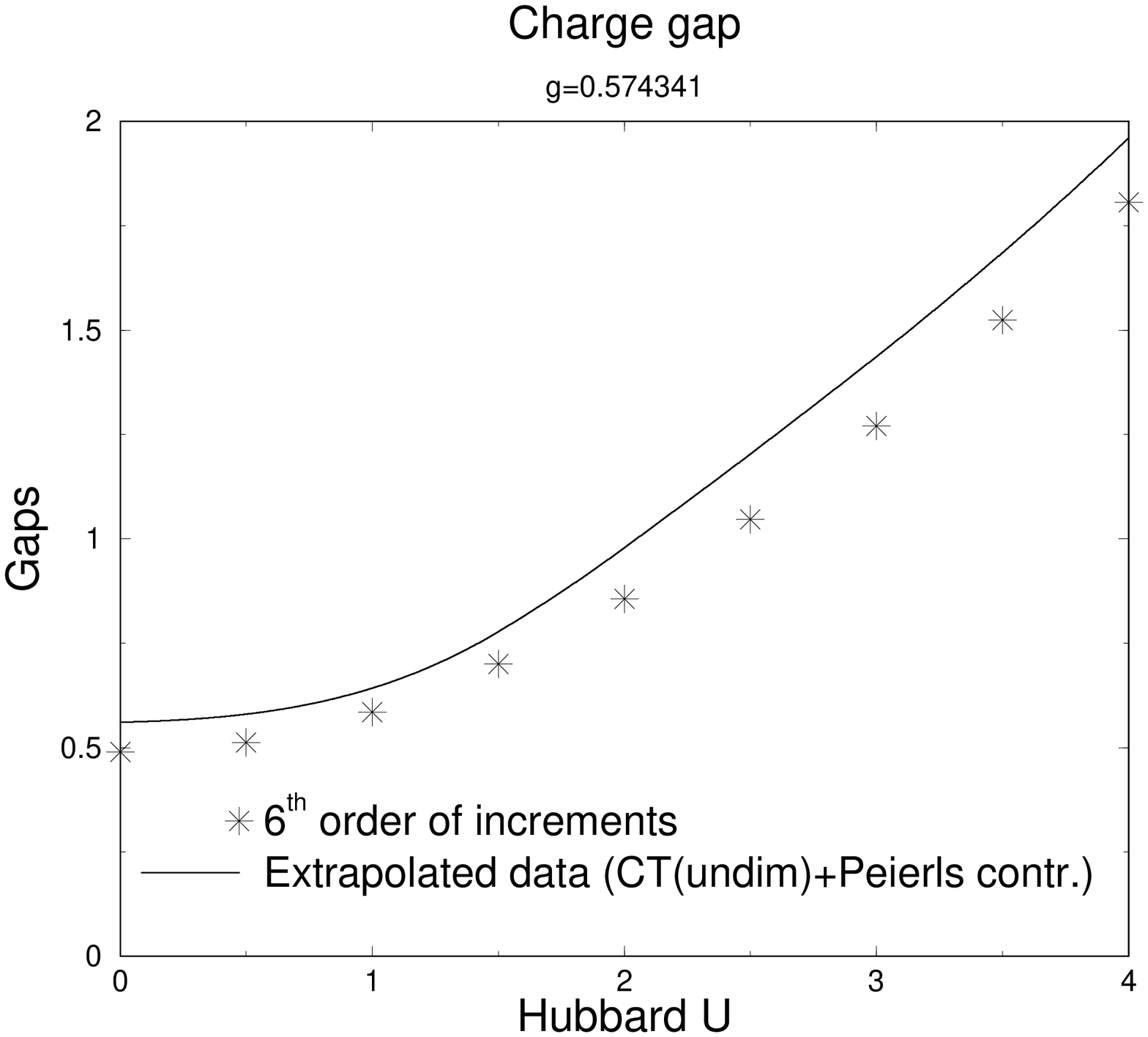,angle=-90,width=7.0cm}
\end{minipage}
\end{center}
\vspace{-0.0cm}
\caption{}
\label{fig5}
\end{minipage}
\end{center}
\end{figure}

\end{document}